\documentclass{elsart}
\usepackage{epsfig}

\def\iproduct#1#2{\big<#1\big|#2\big>}
\def\bra#1{\big<#1\big|}
\def\Dbra#1{\big<\!\big<#1\big|}
\def\ket#1{\big|#1\big>}
\def\Dket#1{\big|#1\big>\!\big>}

\begin{document}

\begin{frontmatter}
\title{Light-Cone Quantization Without Periodic Boundary Conditions}
\author{Masahiro Maeno}
\address{Department of Physics and Earth Sciences, University of the Ryukyus\\
Senbaru 1 Nishihara cho, OKINAWA 903-0213, JAPAN
}
\begin{abstract}
This paper describes a light-cone quantization of a two-dimensional
 massive scalar field without periodic boundary conditions in order to
 make the quantization manifestly consistent to causality. For this
 purpose, the field is decomposed by the Legendre
 polynomials. Creation-annihilation operators for this field are defined
 and the Fock space was constructed.
\end{abstract}
\end{frontmatter}

\section{Introduction}
The subject of this paper is the search for a way to carry out
light-cone quantization with neither periodic nor anti-periodic boundary
conditions.

Recently, much attention has been paid to light-cone quantization.  The
most remarkable feature of light-cone quantization is the simplicity of
its vacuum. Namely, an interacting vacuum is identical to a perturbative
vacuum.  In light-cone quantization, a mode operator which has a
positive (or negative) light-cone momentum($p^+$) is respectively an
annihilation ( or a creation) operator. Due to $p^+$-conservation, all
terms in an interaction Hamiltonian contain at least one annihilation
operator. Therefore, the interaction Hamiltonian annihilates the
vacuum. However, this situation becomes vague by the existence of
so-called `zero-mode'\cite{MaskawaYamawaki}\cite{Ida}\cite{NakaYama}
which has zero light-cone momentum.

The zero mode plays an important role in light-cone quantization. For
example, it can create vacuum structure (see
\cite{SchMC}\cite{SUN}\cite{SchKR}) or control spontaneous symmetry
breaking (see \cite{SSB}\cite{HSWZ}). However, the existence of the zero
mode crucially depends on which boundary condition we choose.

\begin{figure}[h]
\caption{No Boundary Condition}
\epsfbox{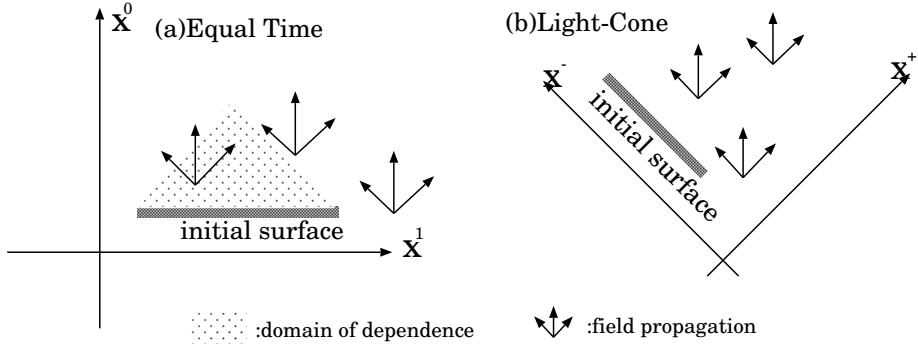}
\label{nobcFig}
\end{figure}
Let us consider the subject of boundary condition from the perspective
of the domain of dependence. Figure \ref{nobcFig} shows domains of
dependence in equal-time and light-cone quantization without boundary
condition.

\begin{figure}[h]
\caption{Periodic Boundary Condition}
\epsfbox{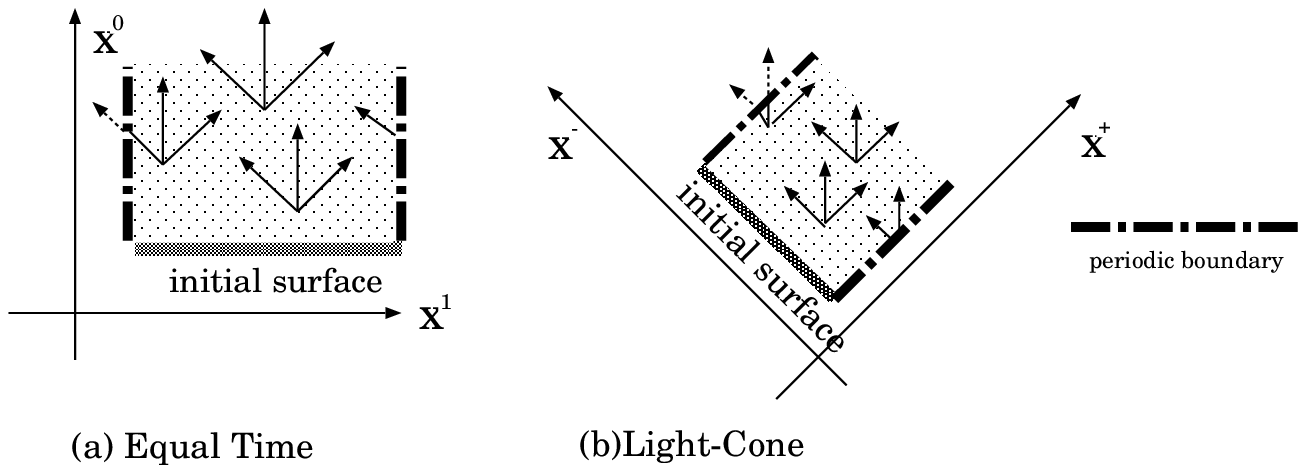}
\label{pbcFig}
\end{figure}

In light-cone quantization, an initial condition is provided on a
constant $x^+$-plane\footnote{Our coordinate convention is $x^\pm =
{1\over\sqrt{2}}\left(x^0\pm x^1\right)$. The domain of $x^-$ is
$[-L,L]$.} (in the equal-time case, a constant $x^0$-plane).  Note that
we have no domain of dependence for light-cone quantization in this
case. This means that we cannot predict the future value of $\phi$ at
any point with initial condition only.  Therefore, we set a periodic (or
anti-periodic) boundary condition $\phi(x^+,L)=\pm\phi(x^+,-L)$ as
Figure \ref{pbcFig}.

In the equal time quantization case, periodic (or anti-periodic)
boundary condition $\phi(x^0,L)=\pm\phi(x^0,-L)$ connects two space-like
separated points. This is physically acceptable, but in the light-cone
quantization case, the periodic boundary condition is slightly unnatural
because it connects two null-like separated points.  This prescription
of periodic boundary condition is probably acceptable for most of the
subjects in light-cone quantization. However, this small problem might
be particularly critical for some cases in which boundary information is
essential.  For this reason, we need to find a method of light-cone
quantization with causally natural boundary condition.

\begin{figure}[h]
\caption{V-Type Boundary Condition}
\epsfbox{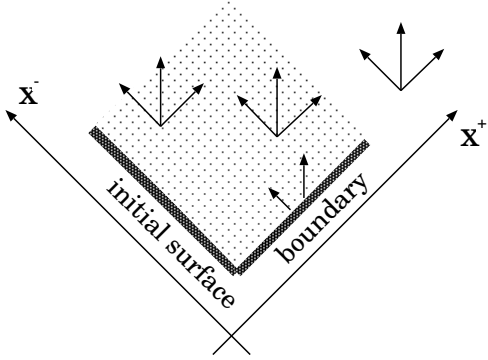}
\label{vbcFig}
\end{figure}

For our purpose, the boundary condition for light-cone quantization
should be set on the V-shaped boundary as shown in Figure
\ref{vbcFig}. We give an initial condition on a constant
$x^+$-plane($\phi(x^+=0,x^-)=\phi_I(x^-)$) and a boundary condition on a
constant $x^-$-plane(
$\partial_+\phi(x^+,x^-=-L)=\partial_+\phi_B(x^+)$). Here, we impose the
boundary condition to $\partial_+\phi(x^+,-L)$ in order to prevent a
possible inconsistency, as the first candidate of boundary condition
$\phi(x^+,x^-=-L)=\phi_B(x^+)$ can conflict with the initial condition
$\phi(x^+=0,x^-)=\phi_I(x^-)$ on a point $x^+=0,x^-=-L$.

\section{Legendre Polynomial Expansion}
To quantize a field, we need to expand the field operator by some
functional basis. Fourier expansion depends strongly on the periodicity;
 therefore, it is not appropriate to our purpose. Let us expand the
field $\phi(x)$ as
\begin{equation}
\phi(x^+,x^-)={1\over\sqrt{2}}\sum_{m=0}^{2N} a_m(x^+) P_m({x^-})
\label{phidef}
\end{equation}
using Legendre polynomials $P_m(x)$. We use a conventional
unit:$L=1$. We can easily recover $L$ by a dimensional analysis whenever
desired. To make the following calculation tractable, we limit the mode
number $m$ from $0$ to $2N$. The number $N$ should be set to infinity
after all calculations are complete.  At the present stage, no boundary
condition is imposed.  We will explain below how information of the
boundary is introduced to this system.

In the following, we represent a function $f(x)$ defined on $[-1,1]$ as
a ket $\ket{f}$. An integral $\int_{-1}^1 f^*(x)g(x) dx$ is represented
by an inner product $\iproduct{f}{g}$. A ket $\ket{x}$ principally
denotes an eigenvector of $x$-representation which satisfies
$\iproduct{x}{x'}=\delta(x-x')$ and $ \int_{-1}^1 dx\ket{x}\bra{x}=1$.

We define a series of kets $\ket{n}(n=0,1,2,\cdots,2N)$ by
\begin{equation}
 {1\over\sqrt{2}}P_n\left({x}\right)=\iproduct{x}{n}
\end{equation}
which satisfies
\begin{eqnarray}
& \iproduct{k}{n}
=\int_{-1}^1 dx \iproduct{k}{x}\iproduct{x}{n}
={1\over2}\int_{-1}^1 dx P_k\left({x}\right)
P_n\left({x}\right)={1\over 2n+1}\delta_{kn}&\\
& \sum_n \ket{n}(2n+1)\bra{n}=1.&
\end{eqnarray}
In the following, any Latin index ($k,l,m,n,\cdots$) repeated in a
product is automatically summed from $0$ to $2N$. Under this notation,
we can write (\ref{phidef}) as
\begin{equation}
\phi(x^+,x^-)=a_m(x^+) \iproduct{x^-}{m}
\end{equation}
Defining a derivative operator $\partial$ as ${\partial\over\partial
x}\bra{x}=\bra{x}\partial$, its derivative can be written as
\begin{equation}
{\partial\over \partial x^-}\phi(x^+,x^-)=a_m(x^+) \bra{x^-}\partial\ket{m}.
\end{equation}

Using well-known formulas for Legendre polynomials, the matrix
elements of the operator $\partial$ are calculated as
\begin{equation}
 \bra{k}\partial\ket{n}=\cases{1 & $k<n, (k+n)$ mod 2 =1 \cr 0 &otherwise}
\end{equation}

The Lagrangian is calculated as
\begin{equation}
 L_{free}= \int_{-1}^1 dx^- \left(\partial_+ \phi \partial_- \phi -
 {m^2\over2}\phi^2 \right)= 
\dot a_k\bra{k}\partial\ket{l}a_l
-{m^2\over 2}a_k\iproduct{k}{l}a_l
\end{equation}
This expression is non-diagonal with respect to $a_n$ because
$\bra{k}\partial\ket{l}$ is a non-diagonal matrix. Although it makes the
following calculation cumbersome, we have to pay this price for avoiding
periodic boundary conditions.

\section{Canonical Quantization}
In this section, we carry out canonical quantization. As explained in
the previous section, we impose the boundary condition $\dot
\phi(x^+,-L) = \dot \phi_B(x^+)$.  For this purpose, we add
\begin{equation}
\sqrt{2}B(\dot \phi_B(x^+)-\dot \phi(x^+,-L))
= \sqrt{2}B\dot \phi_B(x^+) - B (-1)^n \dot a_n
\end{equation}
 to the Lagrangian density.  The new variable $B$ is the Lagrange
 multiplier (a counterpart of the Nakanishi-Lautrup field) and $\phi_B$
 is a classical boundary value of $\phi$.

Conjugate momenta of $a_n$ and $B$ are written as
\begin{eqnarray}
\pi_{k}&=&\bra{k}\partial\ket{l}a_l-(-1)^k B \\
\pi_B&=&0.
\label{priconstraint}
\end{eqnarray}
Because all $\pi$ contain
no time($x^+$)-derivative, these equations must be treated as
constraints.
\begin{eqnarray}
 \varphi_k &\equiv& \pi_k- \bra{k}\partial\ket{l}a_l+(-1)^k B=0\\
 \varphi_B &\equiv& \pi_B=0
\end{eqnarray}
To calculate Dirac brackets, we define new linear combinations of
constraints as
\begin{equation}
\tilde \varphi_{\tilde n}\equiv \varphi_{\tilde n} - (-1)^{\tilde n} \varphi_0
+ \varphi_B \bra{0}\partial\ket{\tilde k}.
\end{equation}
From now on, we use indices with a tilde ($\tilde k,\tilde l,\tilde
n,\cdots$) for numbers running from $1$ to $2N$ (exclude $0$).

The modified constraints are
\begin{eqnarray}
 \tilde \varphi_{\tilde k} &=& \pi_{\tilde k} -(-1)^{\tilde k}\pi_{0}
- \Dbra{\tilde k}\partial\ket{l}a_l + \pi_B \bra{0}\partial\ket{\tilde k}\\
\varphi_0 &=& \pi_0 - \bra{0}\partial\ket{k}a_{k} + B\\
\varphi_B &=& \pi_B
\end{eqnarray}
where we use a notation $\Dbra{n}\equiv \bra{n}-(-1)^n \bra{0}$.
The Poison bracket between these constraints is
\begin{equation}
\bordermatrix{
  &\tilde \varphi_{\tilde l}&\varphi_0 &\varphi_B \cr
  \tilde \varphi_{\tilde k}&\Dbra{\tilde k}\bar \partial\Dket{\tilde l} &0& 0\cr
  \varphi_0&0 &0&-1\cr
  \varphi_B&0 & 1 &0
}
\end{equation}
where $\bar\partial=\partial-\partial^\dagger$. The elements of the
matrix $\Dbra{\tilde k}\bar\partial\Dket{\tilde l}$ and its inverse are
\begin{eqnarray}
 \Dbra{\tilde k}\bar\partial\Dket{\tilde l}&=&\cases{
2 & $\tilde k<\tilde l$, $\tilde k$ is odd, $\tilde l$ is even. \cr 
-2 & $\tilde k>\tilde l$, $\tilde k$ is even, $\tilde l$ is odd. \cr 
0 &otherwise}\\
 \Dbra{\tilde j}\bar\partial^{-1}\Dket{\tilde k}&=& {1\over2}\left(\delta_{\tilde j,\tilde k+1}-\delta_{\tilde j+1,\tilde k}\right).
\end{eqnarray}
All $2N+2$ constraints are second-class. We can calculate the
Dirac brackets as
\begin{equation}
\begin{array}{rl}
 \left\{\psi_1,\psi_2\right\}_{DB}
&=\left\{\psi_1,\psi_2\right\}_{PB}
+\left\{\psi_1,\tilde \varphi_{\tilde k}\right\}_{PB}
 \Dbra{\tilde k}\bar\partial^{-1}\Dket{\tilde l}
\left\{\tilde \varphi_{\tilde l},\psi_2\right\}_{PB}\\
&+\left\{\psi_1,\varphi_0\right\}_{PB}
\left\{\varphi_B,\psi_2\right\}_{PB}
-\left\{\psi_1,\varphi_B\right\}_{PB}
\left\{\varphi_0,\psi_2\right\}_{PB}.
\end{array}
\end{equation}
The results are listed below:
\begin{eqnarray}
\left[a_k,B\right]_{DB}&=&-{1\over2}(\delta_{k,0}+\delta_{k,2N})\\
\left[a_k,a_l\right]_{DB}&=&{1\over2}
\left(\delta_{k+1,l}-\delta_{k,l+1}+\delta_{k,0}\delta_{l,2N}-\delta_{k,2N}\delta_{l,0}
\right)
\end{eqnarray}
Commutation relations are obtained from the above by multiplying
$i\hbar$.

The Hamiltonian has the mass term $H_m$ and the boundary-fixing term
$H_B$ as
\begin{equation}
H_m = {m^2\over2} a_k\iproduct{k}{l} a_l,
~~~
H_B = -\sqrt{2}B\partial_+ \phi_B.
\end{equation}

The time development of the boundary value
$\phi(x^+,-L)={1\over\sqrt{2}}a_n (-1)^n$
is now derived from a commutator with the Hamiltonian as
\begin{equation}
\partial_+ \phi(x^+,-L)=
-i \left[\phi(x^+,-L),H_m + H_B\right]_{DB}
= \partial_+ \phi_B.
\end{equation}
The result is consistent to our boundary condition.

\section{Creation-Annihilation Operators}
In this section, we construct creation-annihilation operators. For
simplicity, we consider $\partial_+\phi_B=0$ case, which is a light-cone
extension of Neumann boundary condition.  Let us define a new operator
$A(\alpha)$ as a linear combination of $a_n$:
\begin{equation}
A(\alpha)= \lambda_n(\alpha)a_n
\label{defofA}
\end{equation}
which satisfies
\begin{equation}
\left[A(\alpha),H_m\right]=\alpha {m^2\over2}A(\alpha)
\label{AandH}
\end{equation}
where $\lambda_n(\alpha)$ are c-number coefficients which depend on
$\alpha$ and $n$.
By the definition, this operator $A(\alpha)$ lowers the energy eigenvalue by
$\alpha{m^2\over2}$. $A(\alpha)$ with positive (or negative)
$\alpha$ are annihilation (or creation) operators, respectively.

In the following, we calculate the coefficients $\lambda_n(\alpha)$.
From (\ref{AandH}), they have to satisfy the following equations:
\begin{eqnarray}
 \lambda_1(\alpha) + \lambda_{2N}(\alpha) &=& i\alpha\lambda_0(\alpha)
\label{zenone}\\
 \lambda_{n+1}(\alpha) - \lambda_{n-1}(\alpha) &=& i\alpha(2n+1)\lambda_n(\alpha)
 ~~~~~~~~~\hbox{for }1 \le n \le 2N-1 \label{zentwo}\\
 -\lambda_0(\alpha) - \lambda_{2N-1}(\alpha) &=& i\alpha(4N+1)\lambda_{2N}(\alpha).\label{zenthree}
\end{eqnarray}
Solving (\ref{zenone}) and (\ref{zentwo}), all $\lambda_n(\alpha)$s can be
written by $\lambda_0(\alpha)$ and $\lambda_{2N}(\alpha)$ as
\begin{equation}
\lambda_m(\alpha)=\sum_{l=0}^m\left({i\alpha\over2}\right)^l{(m+l)!\over l! 
(m-l)!}\times\left\{\begin{array}{cl}-\lambda_{2N}(\alpha)& \hbox{~for odd
$m+l$}\\ \lambda_0(\alpha) &\hbox{~for even $m+l$}\end{array} \right\}
\label{lambdas}
\end{equation}
which can be easily proved by induction.

These equations lead to two expressions for $\lambda_{2N}(\alpha)$.

From (\ref{lambdas}),
 \begin{equation}
 \lambda_{2N}(\alpha)=
 {\sum_{m=0}^N\left({i\alpha\over 2}\right)^{2m}{(2N+2m)!\over (2m)!(2N-2m)!}
 \over 1+\sum_{m=0}^{N-1}\left({i\alpha\over2}\right)^{2m+1}
 {(2N+2m+1)!\over (2m+1)!(2N-2m-1)!}}\times\lambda_0(\alpha).
 \label{lambdaone}
 \end{equation}
From the equation (\ref{zenthree}),
 \begin{equation}
 \lambda_{2N}(\alpha)={
 1+\sum_{m=0}^{N-1}\left({i\alpha\over 2}\right)^{2m+1}{(2N+2m)!\over (2m+1)!(2N-2m-2)!}
 \over
 \sum_{m=0}^{N-1}\left({i\alpha\over 2}\right)^{2m}{(2N+2m-1)!\over (2m)!(2N-2m-1)!}
 -i\alpha(4N+1)
 }\times\lambda_0(\alpha)
 \label{lambdatwo}
 \end{equation}

Agreement of these two expressions implies a condition
for $\alpha$;
\begin{equation}
 \sum_{m=0}^{N}\left({i\alpha\over2}\right)^{2m+1}{(2N+2m+1)!\over
 (2m+1)!(2N-2m)!}=0.  \label{eqAlpha}
\end{equation}
This equation has $2N+1$ solutions as $\alpha=(0,\pm \alpha_1, \pm
\alpha_2,\cdots,\pm\alpha_{N})$.\footnote{Although we do not go into
detail, numerical calculation shows that the solutions of
(\ref{eqAlpha}) are nearly to ${2\over (2n+1)\pi}$. These values are
obtained as a result of $N\to \infty$ limit in the next section.}  
At the present stage, all $\lambda_n(\alpha)$ have a factor
$\lambda_0(\alpha)$. So we can freely set normalization of $\lambda_0(\alpha)$.
 As a convention, we set $\lambda_0(\alpha)$
to be a real number which satisfies
$\lambda_0(\alpha)=\lambda_0(-\alpha)$.  In this convention,
$\lambda_n(-\alpha)=\lambda_n^*(\alpha)$ and
$A(-\alpha)=A^\dagger(\alpha)$. 

Now we introduce a set of vectors in the function space as
\begin{equation}
 \ket{\alpha}=(2n+1)\lambda_n(\alpha)\ket{n}.
\end{equation}
$\ket{\alpha}$ is also a set of linear independent $2N+1$ vectors
$\left\{\ket{\alpha} \big|
\alpha=0,\pm\alpha_1,\pm\alpha_2,\cdots,\pm\alpha_{N}\right\}$.  As
easily confirmed using (\ref{zenone}),(\ref{zentwo}), (\ref{zenthree})
and $\iproduct{m}{n}={1\over2n+1}\delta_{mn}$, $\iproduct\alpha\beta =0$
for $\alpha\ne\beta$. We set normalization of $\lambda_0(\alpha)$ to
$\iproduct\alpha\alpha=1$.  In this convention,
$\lambda_n(\alpha)=\iproduct{\alpha}{n}$. We will use this expression
from now on. We calculate commutation relation of $A(\alpha)$s as
\begin{equation}
\left[A(\alpha),A(\beta)\right]=\alpha{2n+1\over2}\iproduct{\alpha}{n}\iproduct{\beta}n
= {\alpha\over2} \delta_{\alpha+\beta,0}.
\end{equation}

The mass term of the Hamiltonian can be written by $A(\alpha)$ as
\begin{equation}
 H_m= {m^2\over 2}\sum_{\alpha}A(-\alpha)A(\alpha)
\end{equation}
and the field $\phi$ is expanded as
\begin{equation}
 \phi(x^+,x^-) = \sum_{\alpha}e^{-i{\alpha m^2\over2}x^+}\iproduct{x}{\alpha}A(\alpha)
\end{equation}
Here, the function $\iproduct{\alpha}{x}$ plays
 a role of $e^{ip^+ x^-}$
in the periodic boundary condition case.

The Fock space is constructed as
\begin{equation}
 \left[A(-\alpha_1)\right]^{n_1}
 \left[A(-\alpha_2)\right]^{n_2}
\cdots
 \left[A(-\alpha_N)\right]^{n_N}\ket{0}
\times(\hbox{a function of $A(0)$})
\end{equation}
Since the operator $A(0)$ commutes with all other $A(\alpha)$, it does
not belong to the creation-annihilation pair. Hence, we use coordinate
representation for $A(0)$.

We show that we can define creation-annihilation operators even if we do
not use periodic boundary conditions.

\section{$N \to \infty$ Limit}

Now we take a limit $N \to \infty$. First, we divide the equation
(\ref{eqAlpha}) by
$\left({i\alpha\over2}\right)^{2N+1}{(4N+1)!\over(2N+1)!}$ and replace
$m \to N-n$. We get
\begin{equation}
 \sum_{n=0}^{N}\left({i\alpha\over2}\right)^{2n}{(4N-2n+1)!(2N+1)!\over
 (4N+1)!(2n)!(2N-2n+1)!}=0
\end{equation}
Since $\lim_{N \to \infty}{(4N-2n+1)!(2N+1)!\over (4N+1)!(2N-2n+1)!} =
2^{-2n}$, $N\to\infty$ limit of the equation (\ref{eqAlpha}) is
\begin{equation}
\sum_{n=0} {1\over 2n!}\left({i\alpha}\right)^{2n} =0.
\end{equation}
Namely, $\cos\left({1\over\alpha}\right)=0$, it means $\alpha={2\over
(2n+1)\pi}$ where $n=$integer.

Hence, the energy quanta of the present theory are ${m^2L\over(2n+1)\pi}$
(we can know the power of $L$ by a dimensional analysis) in contrast
to one in periodic boundary condition(${m^2L\over 2n\pi}$).  It is
identical to energy quanta in the anti-periodic case.

\section{Conclusion}

In this paper, we quantized two-dimensional massive scalar fields in
light-cone frame. In the process of quantization, we did not impose
periodic boundary condition. Our V-shaped boundary condition is fully
consistent to causality. The Fock space can be constructed in this
formalism.  As a result, we obtained a different spectrum from the case
of periodic boundary condition.

The formalism in this paper is applicable to other theories.  In
particular, an application to gauge theory is interesting.  As pointed
out in many references (see \cite{SchMC} etc.), the light-cone gauge
$A^+=0$ is not appropriate under periodic boundary conditions; the zero
mode of $A^+$ cannot be gauged away by gauge transformation ($A^+ \to
A^+ +\partial^+ \Lambda$). Some vacuum structures, for example,
$\theta$-vacuum in the Schwinger model, come from dynamics of this
residual zero-mode of $A^+$.  However, the light-cone gauge $A^+=0$ is
fully consistent to our boundary condition. The vacuum structure may
come from other sources. This problem is worth attacking.  The most
important question is whether our vacuum is stable even in interacting
theory, and this will remain a problem in the future.

In this paper, we found a new formalism of light-cone quantization.  The
author expects that this new formalism will lead to new features of
light-cone quantization.

\section*{Acknowledgment}
The author would like to thank Dr. Stephen Pinsky and Dr. Motoi
Tachibana for their valuable suggestions and encouragements.

\end{document}